\pdfoutput=1

\documentclass[11pt]{article}

\usepackage[]{ACL2023}

\usepackage{times}
\usepackage{latexsym}

\usepackage[T1]{fontenc}

\usepackage[utf8]{inputenc}

\usepackage{microtype}

\usepackage{inconsolata}

\usepackage{graphicx}
\usepackage{subcaption}
\usepackage[normalem]{ulem}

\usepackage{tikz}
\usetikzlibrary{arrows}
\usepackage{tikz}

\usepackage{tikzscale}
\usetikzlibrary{arrows.meta,
                calc,
                positioning,
                quotes,
                shapes.geometric}

\pgfdeclareshape{datastore}{
  \inheritsavedanchors[from=rectangle]
  \inheritanchorborder[from=rectangle]
  \inheritanchor[from=rectangle]{center}
  \inheritanchor[from=rectangle]{base}
  \inheritanchor[from=rectangle]{north}
  \inheritanchor[from=rectangle]{north east}
  \inheritanchor[from=rectangle]{east}
  \inheritanchor[from=rectangle]{south east}
  \inheritanchor[from=rectangle]{south}
  \inheritanchor[from=rectangle]{south west}
  \inheritanchor[from=rectangle]{west}
  \inheritanchor[from=rectangle]{north west}
  \backgroundpath{
    \southwest \pgf@xa=\pgf@x \pgf@ya=\pgf@y
    \northeast \pgf@xb=\pgf@x \pgf@yb=\pgf@y
    \pgfpathmoveto{\pgfpoint{\pgf@xa}{\pgf@ya}}
    \pgfpathlineto{\pgfpoint{\pgf@xb}{\pgf@ya}}
    \pgfpathmoveto{\pgfpoint{\pgf@xa}{\pgf@yb}}
    \pgfpathlineto{\pgfpoint{\pgf@xb}{\pgf@yb}}
 }
}

\usepackage{adjustbox}

\title{Redesigning Electronic Health Record Systems to Support Developing Countries}

  \author{Jean Marie Tshimula,$^{1,2,4}$ D'Jeff K. Nkashama,$^{1,2}$ Kalonji Kalala,$^{1,3}$ Maximilien V. Dialufuma,$^{1,7}$ \\ {\bf Mbuyi Mukendi Didier,$^{1,6,8}$ Hugues Kanda,$^{1}$  Jean Tshibangu Muabila,$^{1,5}$ Christian N. Mayemba$^{1,9}$} \\
  $^{1}$Groupe de Recherche de Prospection et Valorisation des Données (Greprovad) $^{2}$Université de Sherbrooke \\ $^{3}$EECS, University of Ottawa $^{4}$Université TÉLUQ  $^{5}$LISV-UVSQ, Université Paris-Saclay $^{6}$University of \\ Kinshasa $^{7}$Montreal Behavioural Medicine Centre, CIUSSS-NIM/UQAM/Concordia $^{8}$Biomedical Research \\ Unit, Hospital Monkole, Kinshasa $^{9}$Department of Anatomical Pathology, University of Kinshasa \\
  {\tt \iffalse \{\nkad2101, kabj2801\}\fi contact@greprovad.org} \\ }

\date{}

\begin{document}
\maketitle
\begin{abstract}
Electronic Health Record (EHR) has become an essential tool in the healthcare ecosystem, providing authorized clinicians with patients' health-related information for better treatment. While most developed countries are taking advantage of EHRs to improve their healthcare system, it remains challenging in developing countries to support clinical decision-making and public health {\color{black}using a computerized patient healthcare information system}. This paper proposes a novel EHR architecture suitable for developing countries---an architecture that fosters inclusion and provides solutions tailored to all social classes and socioeconomic statuses. Our architecture foresees an internet-free (offline) solution to allow medical transactions between healthcare organizations, and the storage of EHRs in geographically underserved and rural areas. Moreover, we discuss how artificial intelligence can leverage anonymous health-related information to enable better public health policy and surveillance. 
\end{abstract}

\section{Introduction} 

Electronic health record (EHR) systems provide a secure, integrated collection of patient and population electronically-stored health information in a digital format~\cite{Odekunle01,Kukafka01,Akanbi0001,Adetoyi0001,Kavuma0001,Kohli0002}; it provides a comprehensive digital view of a patient’s health history with the goals of eliminating legibility problems with handwritten records; enabling remote access of health records; facilitating intervention earlier in the course of the disease, patient care, and outcomes; increasing efficiency and lowering costs; and ameliorating billing procedures~\cite{Schmitt001,Erstad001}. The potential benefits of EHR systems have enabled its wide adoption in developed and some emerging countries~\cite{Black0001}.

While most developed countries are taking advantage of EHRs to improve their healthcare system, it remains challenging in developing countries to support clinical decision-making and public health {\color{black}using a computerized patient healthcare information system}. Some developing countries including sub-Saharan Africa still predominantly use paper-based systems in healthcare delivery, instead of computerized patient management systems~\cite{Odekunle01,Akanbi0001,Adetoyi0001,Kavuma0001,Kohli0002}. The lack of an EHR system may lead to issues in managing patient health data to improve the quality of patient care and safety through decision support to clinicians. For instance, patient \texttt{P} lives in city \texttt{X}, travels to city \texttt{Y} in the same country and falls sick during her stay. Since clinician \texttt{C} in \texttt{Y} does not have more health data about patient \texttt{P}, (i) treatment options provided to \texttt{P} could cause some important problems involving past health issues and (ii) prescription drugs delivered to \texttt{P} could ignore her medical history. Medication errors can result in a substantial economic burden on patients, side effects, and irreversible consequences; there is a huge spectrum of medication errors. Some errors may be minors and others may lead to adverse events causing complications and higher mortality~\cite{Bates01,Forster0001}. However, EHR systems can potentially reduce prescription errors and adverse drug interactions~\cite{Chaudhry01} and make available medical history data during emergency care~\cite{Stiell01}. This data provides vital medical history details and gives more options to clinicians to decide which treatment best corresponds to the problem and when it should be administered. We pose the question: \textit{``How could we replace paper-based systems with EHR systems in the context of developing countries?''}. A study identified some factors hindering the widespread adoption of EHR systems in developing countries. The identified factors include but are not limited to high cost of procurement and maintenance, poor electricity supply and internet connectivity~\cite{Odekunle01}. This paper therefore proposes an EHR architecture that addresses the previously mentioned factors.

\begin{figure}[t!]
    \centering
    \small 
\begin{center}    
   \begin{tikzpicture}[
  font=\sffamily,
  every matrix/.style={ampersand replacement=\&,column sep=.75cm,row sep=0.8cm},
  source/.style={draw,thick,rounded corners,fill=yellow!20,inner sep=.3cm},
  process/.style={draw,thick,circle,fill=blue!20},
  sink/.style={source,fill=green!20},
  datastore/.style={draw,very thick,shape=datastore,inner sep=.5cm},
  datastore2/.style={draw,very thick,shape=datastore,inner sep=.5cm},
  datastore3/.style={draw,very thick,shape=datastore,inner sep=.5cm}
  dots/.style={gray,scale=2},
  to/.style={->,>=stealth',shorten >=1pt,semithick,font=\sffamily\footnotesize},
  every node/.style={align=center},
  persistence/.style = {cylinder, draw, shape border rotate=90, minimum height=11mm, minimum width=14mm}
  ]

  \matrix{
    
    \& ;
    \& ;
    \node[source] (hisparcbox) {EHR}; \& \\ 
    
    \& \node[sink] (datastore) {Web\\ system}; 
    \& \node[sink] (datastore2) {Mobile\\ system}; 
    \& \node[sink] (datastore3) {USSD\\ system};  \\
  };

  \draw[>=triangle 45, <->] (hisparcbox) to (datastore2);

  \node (dts) [persistence, below=of datastore2, label=below:{EHR Database (DB)}]  {};
  
  \draw[>=triangle 45, <->] (datastore2) to node {\quad\quad\textcircled{2}} (dts) ;
  
  \draw[>=triangle 45, <->] (dts.west)  -| node[anchor=north,pos=.5]{\textcircled{1}}  (datastore.south);

\draw[>=triangle 45, <->] (dts.east) -| (datastore3.south) node[anchor=north,pos=0.5]{\textcircled{3}} ;

\draw[>=triangle 45, <->] (hisparcbox) -| node[anchor=north]{} (datastore);

\draw[>=triangle 45, <->] (hisparcbox) -| (datastore3);

\end{tikzpicture}
\end{center}  
    \caption{Architectural diagram of an EHR system tailored to the context of developing countries.}
    \label{archglob}
    \vspace{-5mm}
\end{figure}
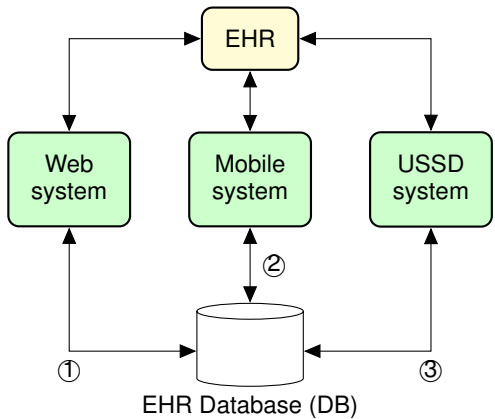

We believe that the implementation of EHR systems in the style of industrialized countries may fail to function and provide solutions in the context of developing countries. To implement an EHR system in developing countries, besides the aforementioned issues, we also address the issues related to social inclusion, discrimination and socioeconomic status in healthcare. Everyone qualifies for health monitoring regardless of personal income, or standard of living. We propose a straightforward architecture to implement an EHR system that fosters inclusion and provides solutions tailored to all social classes. The proposed architecture takes into consideration internet coverage, electricity, and infrastructure issues and foresees alternative solutions to skirt these issues. More interestingly, our architecture proposes an internet-free alternative (an offline solution) to allow medical transactions within hospitals and clinics and the storage of EHRs in geographically underserved and rural areas. Note that the offline solution does not require relatively expensive terminals (such as computers, tablets, and smartphones) to establish connections between healthcare organizations. The motivation behind this solution is to bridge inequalities in healthcare and allow healthcare organizations with limited means to access EHR systems with any type of mobile phone that they possess. Additionally, the proposed architecture foresees the utilization of artificial intelligence to enable better public health policy and surveillance in (i) monitoring patterns suggesting disease outbreaks, (ii) predicting disease based on symptoms, medical records, and treatments over time, and (iii) providing drug recommendations.

The rest of this paper is organized as follows. A brief outline of some related work is given in \S\ref{r_w}. Section \ref{p_a} describes the proposed architecture. We discuss the scope of the proposed architecture, challenges, and opportunities in \S\ref{discussion}. We describe ethical considerations in \S\ref{e_c}. Finally, we conclude and present future directions in \S\ref{c_f}.



\section{Related work} \label{r_w} 

Researchers investigated qualitative and quantitative methods of storing patient data and reported that an electronic storage and indexing system are a more suitable method for administering medical records~\cite{Dieter0002}. In order to manage patient data, many studies addressed the problem of the implementation and adoption of EHR systems in the context of developing countries~\cite{Adetoyi0001,Odekunle01,Akanbi0001,Kavuma0001,Akwaowo0002,Sood0002,Fraser0002,Syzdykova0002,Kamadjeu0002,Sabi0002}. For instance, Adetoyi and Raji~\shortcite{Adetoyi0001} proposed a design framework for inclusion of EHRs in medical informatics of sub-Saharan Africa; and Kamadjeu et al.~\shortcite{Kamadjeu0002} experimented with the use of an EHR system in urban primary health care practice in Cameroon. Jawhari et al.~\shortcite{Jawhari0002} examined EHR deployments in sub-Saharan African slums by considering the systems, people, processes, and product factors that endorse a crucial involvement in the fate of its implementation, also equating difficulties in knowledge and learning opportunities for EHR use in resource-constrained settings. On similar lines, Kavuma~\shortcite{Kavuma0001} evaluated the implementation of electronic medical record systems in sub-Saharan Africa and then assessed their usability based on a defined set of metrics. While the EHR systems proposed in \cite{Adetoyi0001,Kamadjeu0002} have implemented good strategies to permit healthcare personnel to quickly access patient data to support healthcare delivery, many factors still affect their adoption. The studies in~\cite{Odekunle01,Sabi0002,Akwaowo0002} highlighted some factors hindering the facilitation of broad adoption of EHR systems in sub-Saharan Africa, including the lack of infrastructure, electricity outages, and internet coverage issues~\cite{Odekunle01}. In this paper, we introduce an architecture that addresses these factors and proposes an internet-free solution for accessing an EHR system with relatively minor dependence on the previously mentioned factors. The rationale behind this is to facilitate the use of EHRs in healthcare delivery, make the EHR system accessible to everyone without exception, including slums and rural areas and regardless of socioeconomic status, and combat inequalities in healthcare.

\begin{figure}[t!]
\fontsize{6pt}{6pt}\selectfont
\begin{subfigure}{.5\textwidth}
\begin{center}
\begin{tikzpicture}[
  font=\sffamily,
  every matrix/.style={ampersand replacement=\&,column sep=1.5cm,row sep=1.5cm},
  source/.style={draw,thick,rounded corners,fill=yellow!20,inner sep=.3cm},
  process/.style={draw,thick,circle,fill=blue!20},
  sink/.style={source,fill=green!20},
  datastore/.style={draw,very thick,shape=datastore,inner sep=.3cm},
  dots/.style={gray,scale=2},
  to/.style={->,>=stealth',shorten >=1pt,semithick,font=\sffamily\footnotesize},
  every node/.style={align=center},
  persistence/.style = {cylinder, draw, shape border rotate=90, minimum height=11mm, minimum width=14mm}
  ]

  \matrix{
    \node[source] (hisparcbox) {Web system};
      \& \node[process] (daq) {Internet};  \\
  };

  \draw[to] (hisparcbox) to[bend right=50] node[midway,below] {\fontsize{6pt}{6pt}\selectfont send request}
      node[midway,below] {}  (daq);
      
  \draw[to] (daq) to node[midway,below] {\quad}
      node[midway,above] {\fontsize{6pt}{6pt}\selectfont get response} (hisparcbox);
      
\node (dts) [persistence, right=of daq, label=below:DB]  {};

 \draw[to] (daq) to  node[midway,below] {\quad \\\ {\fontsize{6pt}{6pt}\selectfont data request}}
      node[midway,below] {}  (dts);
      
  \draw[to] (dts) to[bend right=50] node[midway,above] {}
      node[midway,above] {\fontsize{6pt}{6pt}\selectfont fetch data} (daq);

\end{tikzpicture}
\end{center}
\caption{Web application: an internet-dependent system}
\label{archweb}
\end{subfigure} \vfill
\begin{subfigure}{.5\textwidth}

\begin{center}
\begin{tikzpicture}[
  font=\sffamily,
  every matrix/.style={ampersand replacement=\&,column sep=1.5cm,row sep=1.5cm},
  source/.style={draw,thick,rounded corners,fill=yellow!20,inner sep=.3cm},
  process/.style={draw,thick,circle,fill=blue!20},
  sink/.style={source,fill=green!20},
  datastore/.style={draw,very thick,shape=datastore,inner sep=.3cm},
  dots/.style={gray,scale=2},
  to/.style={->,>=stealth',shorten >=1pt,semithick,font=\sffamily\footnotesize},
  every node/.style={align=center},
  persistence/.style = {cylinder, draw, shape border rotate=90, minimum height=11mm, minimum width=14mm}
  ]

  \matrix{
    \node[source] (hisparcbox) {Mobile system};
      \& \node[process] (daq) {Internet};  \\
  };

  \draw[to] (hisparcbox) to  node[midway,above] {}
      node[midway,below] {\fontsize{6pt}{6pt}\selectfont send request}  (daq);
      
  \draw[to] (daq) to[bend right=50] node[midway,above] {}
      node[midway,above] {\fontsize{6pt}{6pt}\selectfont get response} (hisparcbox);
      
\node (dts) [persistence, below=of daq, label=below:DB]  {};

 \draw[to] (daq) to  node[midway,left] {\fontsize{6pt}{6pt}\selectfont data request}
      node[midway,below] {}  (dts);
      
  \draw[to] (dts) to[bend right=30] node[midway,above] {}
      node[midway,right] {\fontsize{6pt}{6pt}\selectfont fetch data} (daq);
      
\node (dts2) [persistence, below=of hisparcbox, label=below:{DB Lite}]  {};

 \draw[to] (hisparcbox) to [bend right=30] node[midway,left] {\fontsize{6pt}{6pt}\selectfont data request}
      node[midway,below] {}  (dts2);
      
  \draw[to] (dts2) to node[midway,above] {}
      node[midway,right] {\fontsize{6pt}{6pt}\selectfont fetch data} (hisparcbox);

  \draw[dashed,to,->] (dts2) to [bend right=50] node[midway,below] {sync} (dts);

\end{tikzpicture}
\end{center}

\caption{Mobile-based EHR system working on offline and \\ online settings}
\label{archmob}
\end{subfigure} \vfill
\begin{subfigure}{.5\textwidth}
\begin{center}
\begin{tikzpicture}[
  font=\sffamily,
  every matrix/.style={ampersand replacement=\&,column sep=1.5cm,row sep=1.5cm},
  source/.style={draw,thick,rounded corners,fill=yellow!20,inner sep=.3cm},
  process/.style={draw,thick,circle,fill=blue!20},
  sink/.style={source,fill=green!20},
  datastore/.style={draw,very thick,shape=datastore,inner sep=.3cm},
  dots/.style={gray,scale=2},
  to/.style={->,>=stealth',shorten >=1pt,semithick,font=\sffamily\footnotesize},
  every node/.style={align=center},
  persistence/.style = {cylinder, draw, shape border rotate=90, minimum height=11mm, minimum width=14mm}
  ]

  \matrix{
    \node[source] (hisparcbox) {USSD};
    \& \node[sink] (datastore) {USSD\\Gateway}; \& \\ 
      \& \node[process] (daq) {Internet};  
       \\
  };

  \draw[to] (hisparcbox) to  node[midway,above] {\fontsize{6pt}{6pt}\selectfont user dials}
      node[midway,below] {}  (datastore);
      
  \draw[to] (datastore) to[bend right=50] node[midway,above] {}
      node[midway,above] {\fontsize{6pt}{6pt}\selectfont get response} (hisparcbox);
     
  \draw[to] (datastore) to[bend right=50] node[midway,left] {\fontsize{6pt}{6pt}\selectfont send request}
      node[midway,below] {}  (daq);
      
  \draw[to] (daq) to node[midway,above] {}
      node[midway,right] {\fontsize{6pt}{6pt}\selectfont get response} (datastore);     
     
\node (dts) [persistence, right=of daq, label=below:DB]  {};

 \draw[to] (daq) to  node[midway,below] {\fontsize{6pt}{6pt}\selectfont data request}
      node[midway,below] {}  (dts);
      
  \draw[to] (dts) to[bend right=50] node[midway,above] {}
      node[midway,above] {\fontsize{6pt}{6pt}\selectfont fetch data} (daq);

\end{tikzpicture}
\end{center}
\caption{EHR system using USSD}
\label{archussd}
\end{subfigure}
\vspace{-3mm}
\caption{Illustration of the proposed architecture at the module level}
\label{fig:fig}
\vspace{-3mm}
\end{figure}
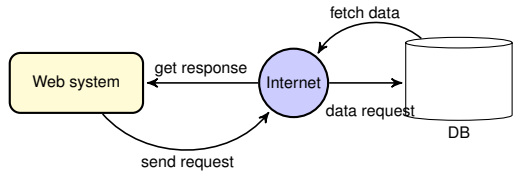
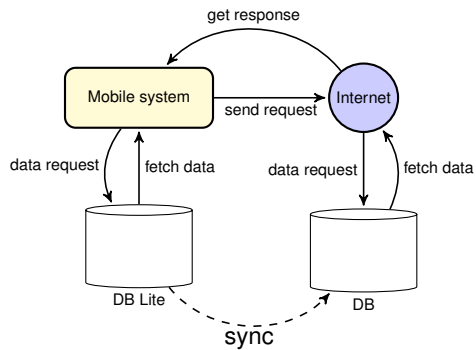
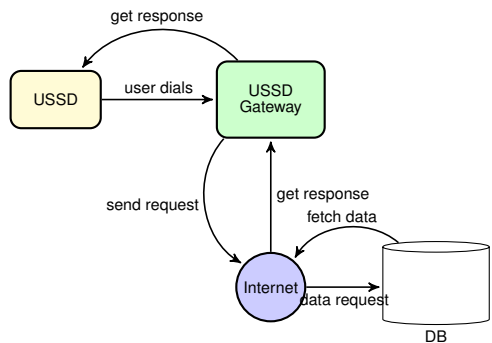

The geographical restrictions of the internet represent an important challenge for the development of Africa~\cite{Counted0002}. Research discovered that demographic and socioeconomic factors as well as complementary infrastructure are also important factors in internet adoption~\cite{Castelan0002}. Usually, individuals utilize mobile broadband internet to access the internet in developing countries, in order to get maternal health support, detect fake drugs and access a digital health-financing platform~\cite{Holst000002}. In healthcare settings, the internet can help clinicians rapidly access patient data and identify suitable treatment plans. Since the lack of infrastructure slows down EHR adoption, our architecture proposes a simple solution that bridges inequalities in healthcare. The particularity of our solution is that it utilizes technology, such as short code\footnote{Understanding the Common Short Code: Its Use, Administration, and Tactical Elements: \\{https://identitypraxis.com/2006/09/01/understanding-the-common-short-code-its-use-administration-and-tactical-elements}. Accessed 2 January 2023} and Unstructured Supplementary Service Data (USSD),\footnote{USSD (Unstructured Supplementary Service Data): \\ {https://www.techtarget.com/searchnetworking/definition/USSD}. Accessed 2 January 2023} to access EHR systems from mobile phones. USSD is a communication protocol used by mobile devices to communicate with a network service provider. USSD can be used to access various services, such as checking account balances or subscribing to a service. It is possible to use USSD to access electronic health records, but this typically involves the creation of a system in which healthcare providers can input and update information into the EHR system using USSD, and access patient records by sending a USSD request to the system. This could be useful in situations where patients or healthcare providers do not have access to a computer or smartphone, or where internet connectivity is unreliable.
 
\section{Proposed architecture} \label{p_a} 

We propose an EHR system management architecture suitable for developing countries. The proposed architecture improves the accessibility of patient data by clinicians in urban, geographically underserved and rural areas. Figure \ref{archglob} illustrates the proposed architecture, which includes a centralized database (EHR Database), and a web-based, mobile-based, and USSD-based EHR system. This section presents each sub-system (component) in the architecture and discusses its specific role and constraints to work properly.     








\subsection{Module 1: Web-based EHR system}\label{mod_webapp}

The web-based EHR system (WES) allows clinicians to access patient medical records via a browser of any device, including a computer and smartphone. As shown in Figure \ref{archweb}, WES could be hosted on a server with uninterrupted internet access. Note that communication between the WES and EHR database also goes over the internet. Interactions start from the WES; for instance, if a clinician wants to access a patient's prescription records. The clinician should first provide credentials for authentication purposes. Once the clinician successfully logs in, the system sends her request to the web server over the internet, and then the web server queries the EHR database to obtain the requested data from WES. Ultimately, the web server responds to the web application with the requested information. One of the advantages of WES is it provides a more friendly interface. It can display different types of data, including images such as medical scans. However, we noticed the architecture of WES depends strongly on the internet as its 
sole communication channel. Consequently, WES can solely be used at hospitals that can afford computers, internet, and servers. In the context of developing countries,  WES seems to be an inadequate solution because it constantly requires the internet and electricity.   

\subsection{Module 2: Mobile-based EHR system } \label{mod_mobapp}

Unlike WES, \if where the web application can be accessed from any device with a browser via the internet,\fi the mobile-based EHR system (MES) allows access to EHRs only on mobile devices. Figure \ref{archmob} shows the functioning of MES. Interactions in MES are slightly similar to WES; the major difference resides in the mobility and capability of MES to work offline. The instability (or lack) of the internet and untimely power cuts in developing countries handicap hospitals to operate normally. Therefore, we propose that MES embeds a lite database called \textit{DB lite}. In the absence of the internet, EHRs could be stored in this database locally. A syncing process is foreseen to transfer locally stored data to the centralized database when the internet connection is established; the syncing process aims to ensure that both DB lite and the centralized database are up to date. Since MES requires smartphones and partially the internet to work properly, it could be challenging for healthcare organizations that are unable to afford smartphones and internet subscriptions. 




\subsection{Module 3: USSD-based EHR system} \label{mod_ussdapp}


USSD (Unstructured Supplementary Service Data) is a communication protocol used by GSM cellular networks to send text messages between a mobile phone and an application program in the network. It allows users to access various services, such as banking and information services, by dialing a code on their phone's keypad and following the prompts. USSD messages are transmitted over the same channels as voice calls, but they do not require a dedicated connection to be established, as is the case with SMS (Short Message Service) texts. This makes USSD a faster and more efficient way to send text-based data between a mobile phone and a server~\cite{Zhou000005,Lakshmi000005}.

An EHR architecture using USSD could enable healthcare providers to access and manage patient data and interact with the healthcare system, using simple text-based commands sent via USSD, even in areas with limited internet connectivity (see Figure \ref{archussd}). (i) The USSD gateway acts as the interface between the mobile network and the EHR system. It is responsible for receiving USSD requests from mobile phones, parsing the requests, and sending them to the EHR system for processing. (ii) The EHR server stores and manages the electronic health records of patients. It receives requests from the USSD gateway, processes them, and sends back the appropriate response. (iii) The EHR database stores electronic health records and other relevant data, such as patient demographics, medications, allergies, and medical history. (iv) The USSD menu system provides the interactive menu system that mobile users interact with when using the EHR system via USSD. It allows users to navigate through the different options and select the appropriate one to perform a specific task, such as viewing their medical history or requesting a prescription refill. (v) The security layer is responsible for ensuring the confidentiality and integrity of the data transmitted between mobile phones and the EHR system. It may include measures such as encryption and authentication.

One of the advantages of the USSD-based EHR system (UES) over WES and MES is that it can be used on any type of mobile phone, even feature phones that do not have internet access; this means that it can potentially be accessed by a wider range of users, including those in rural or low-income areas where internet access may be limited. USSD is a very simple and user-friendly technology, requiring only a basic understanding of how to use a phone keypad; this makes it easy for users to navigate and use the EHR system, even if they are not technically savvy. USSD communication is conducted over the airwaves and is not stored on the device, which means that it is relatively less vulnerable to hacking or data breaches compared to WES and MES. We believe that UES can be a suitable solution for facilitating the utilization of the EHR system in geographically underserved and rural areas, as it requires fewer resources and infrastructure.

\section{Discussion} \label{discussion}

This paper presents a novel EHR architecture suitable for developing countries---an architecture organized into three sub-systems. We discuss the benefits of each sub-system and show how the proposed architecture fosters inclusion and provides solutions adapted to all social classes and socioeconomic statuses. We show that WES and MES depend on the internet and infrastructure such  as computers, smartphones, and servers, even if MES has an offline option that allows it to store EHR data locally, and further syncs this data to the centralized database. A limitation of MES is that the centralized database gets completely updated only when locally stored EHRs are synced. This could be critical in certain scenarios. {\it For instance, suppose that patient \texttt{P} was admitted to hospital \texttt{H1} and her EHRs were stored locally in a device at \texttt{H1}. For medical reasons, hospital \texttt{H1} decides to transfer \texttt{P} to hospital \texttt{H2} for more highly specialized care. Hospital \texttt{H2} could encounter difficulty accessing the recent EHRs of patient \texttt{P} and providing a medical intervention because of the sync issue.} While WES and MES provide straightforward scenarios, we demonstrated that these systems are inadequate for developing countries because of untimely power cuts and internet issues. We argue that UES bridges inequalities in healthcare in areas with limited internet or technological infrastructure to access their health records and other healthcare information. Similar to WES and MES, UES can help healthcare providers make more informed decisions about treatment, reducing the risk of errors or misdiagnosis. One of the advantages of UES is it provides the possibility to access health records at a lower cost and can improve the quality of care for underserved populations by providing healthcare providers with access to a patient's complete medical history. UES can help reduce healthcare costs for underserved populations by enabling the sharing of medical records and other healthcare information between healthcare providers. This can help prevent duplication of tests and other unnecessary procedures, which can be particularly important for patients who may not have the resources to pay for multiple visits or procedures. {\color{black}There are also some potential drawbacks to using USSD for EHR systems, including the limited amount of data that can be transmitted using USSD and the lack of support for multimedia content.}

Our architecture proposes EHR systems that respond to the limitations of~\cite{Adetoyi0001,Kamadjeu0002,Jawhari0002}, foster social inclusion, and can facilitate EHR adoption in developing countries. Many healthcare providers in developing countries may not be familiar with using EHR systems, so providing training and support can help ensure that they are able to use the systems effectively~\cite{Akwaowo0002,Odekunle01,Fraser0002}.

Beyond EHR data storage and manipulation, we can utilize artificial intelligence (AI) to analyze EHR data to enable public health policy and surveillance in a number of ways. One potential use of AI is the analysis of large amounts of data contained in EHRs to identify patterns and trends in the data that may be relevant to public health. This can help public health officials identify potential health threats and take appropriate action to prevent or mitigate them. Another use of AI in the context of public health surveillance is the ability to predict and prevent outbreaks of infectious diseases by analyzing EHR data along with other factors such as population density, travel patterns, and weather conditions~\cite{Yang1000005,Schwartz1000005,Solares1000005,Wong333000005}. By identifying clusters of patients with similar symptoms or diagnoses, AI models can help public health officials anticipate where and when outbreaks are likely to occur and take steps to prevent them. In addition, AI can be used to optimize the allocation of limited public health resources, such as vaccines and medications, by analyzing EHR data to identify which patients are most in need of these resources. Overall, the use of AI in EHRs can help to improve the effectiveness of public health policy and surveillance efforts, particularly in developing countries where access to data and resources may be limited.



\section{Ethical considerations} \label{e_c}
There are important ethical considerations to cover in regard to EHR systems. It is important to ensure that patient information is always kept confidential and secure. This means that access to EHRs should be restricted to authorized healthcare personnel only and that measures such as encryption and secure authentication should be used to protect patient data. Patients have a right to privacy; i.e., personal information should not be shared without the patient's consent and that measures should be in place to prevent unauthorized access to patient data.  Patients have the right to know who has access to their personal data, and to control who can view and use it~\cite{Ozair000005,Genes000005}.  

{\color{black}Note that EHR data can serve to the improvement of public health policies and allow governments or health organizations to make timely decisions. For such a purpose, patient data should be anonymous and aggregated. It is critical to ensure that no reverse engineering could result in patient personal information. For data analysis by geographical zones, we suggest the withdrawal of geographical zones with very low population density in such analysis in order to avoid the disclosure of personal information from aggregates. }   




\section{Conclusion} \label{c_f}

This paper proposes a novel EHR architecture tailored to the context of developing countries. The proposed architecture considers issues related to the internet and electricity and the lack of infrastructure in developing countries and provides solutions adapted to geographically underserved and rural areas. We show how this architecture fosters social inclusion and discuss how the use of AI, on data stemming from the proposed architecture, can help to improve the effectiveness of public health policy and surveillance efforts in developing countries. Additionally, we discuss a few measures and ethical considerations that should be taken while manipulating patient data. In the future, we would like to build AI models that use metadata-induced contrastive learning to (i) provide drug recommendations within an EHR system and (ii) learn patient representations from EHR data to predict dangerous cases of polypharmacy usage and discover sociodemographic biases in the outcomes of polypharmacy usage. 

\section*{Acknowledgments}
The authors thank Moise Mbikayi, René Manassé Galekwa, Senghor Abraham Gihonia, and Cady Nyombe Gbomosa for helpful discussions and comments on early drafts. 

\bibliography{custom}

\begin{thebibliography}{32}
\expandafter\ifx\csname natexlab\endcsname\relax\def\natexlab#1{#1}\fi

\bibitem[{Adetoyi and Raji(2020)}]{Adetoyi0001}
Oluyemi~E. Adetoyi and Olayanju~A. Raji. 2020.
\newblock Electronic health record design for inclusion in sub-saharan africa
  medical record informatics.
\newblock \emph{Scientific African,}, 7:e00304.

\bibitem[{Akanbi et~al.(2012)Akanbi, Ocheke, Agaba, Daniyam, Agaba, Okeke, and
  Ukoli}]{Akanbi0001}
Maxwell~O. Akanbi, Amaka~N. Ocheke, Patricia~A. Agaba, Comfort~A. Daniyam,
  Emmanuel~I. Agaba, Edith~N. Okeke, and Christiana~O. Ukoli. 2012.
\newblock Use of electronic health records in sub-saharan africa: Progress and
  challenges.
\newblock \emph{J Med Trop,}, 14(1):1--6.

\bibitem[{Akwaowo et~al.(2022)Akwaowo, Sabi, Ekpenyong, Isiguzo, Andem, Maduka,
  Dan, Umoh, Ekpin, and Uzoka}]{Akwaowo0002}
Christie~Divine Akwaowo, Humphrey~Muki Sabi, Nnette Ekpenyong, Chimaobi~M.
  Isiguzo, Nene~Francis Andem, Omosivie Maduka, Emem Dan, Edidiong Umoh,
  Victory Ekpin, and Faith-Michael Uzoka. 2022.
\newblock Adoption of electronic medical records in developing countries —a
  multi-state study of the nigerian healthcare system.
\newblock \emph{Frontiers in Digital Health,}, (4):1017231.

\bibitem[{Ayala~Solares et~al.(2020)Ayala~Solares, Diletta~Raimondi, Zhu,
  Rahimian, Canoy, Tran, Pinho~Gomes, Payberah, Zottoli, Nazarzadeh, Conrad,
  Rahimi, and Salimi-Khorshidi}]{Solares1000005}
Jose~Roberto Ayala~Solares, Francesca~Elisa Diletta~Raimondi, Yajie Zhu,
  Fatemeh Rahimian, Dexter Canoy, Jenny Tran, Ana~Catarina Pinho~Gomes, Amir~H
  Payberah, Mariagrazia Zottoli, Milad Nazarzadeh, Nathalie Conrad, Kazem
  Rahimi, and Gholamreza Salimi-Khorshidi. 2020.
\newblock Deep learning for electronic health records: A comparative review of
  multiple deep neural architectures.
\newblock \emph{J Biomed Inform}, 101:103337.

\bibitem[{Bates and Slight(2014)}]{Bates01}
David~W. Bates and Sarah~P. Slight. 2014.
\newblock Medication errors: What is their impact?
\newblock \emph{Mayo Clin Proc.,}, 89(8):1027--102.

\bibitem[{Black et~al.(2011)Black, Car, Pagliari, Anandan, Cresswell, Bokun,
  McKinstry, Procter, Majeed, and Sheikh}]{Black0001}
Ashly~D. Black, Josip Car, Claudia Pagliari, Chantelle Anandan, Kathrin
  Cresswell, Tomislav Bokun, Brian McKinstry, Rob Procter, Azeem Majeed, and
  Aziz Sheikh. 2011.
\newblock The impact of ehealth on the quality and safety of health care: a
  systematic overview.
\newblock \emph{PLoS Med,}, 8(1):e1000387.

\bibitem[{Chaudhry et~al.(2006)Chaudhry, Wang, Wu, Maglione, Mojica, Roth,
  Morton, and Shekelle}]{Chaudhry01}
Basit Chaudhry, Jerome Wang, Shinyi Wu, Margaret Maglione, Walter Mojica,
  Elizabeth Roth, Sally~C. Morton, and Paul~G. Shekelle. 2006.
\newblock Systematic review: impact of health information technology on
  quality, efficiency, and costs of medical care.
\newblock \emph{Ann Intern Med.,}, 144(10):742--52.

\bibitem[{Counted and Arawole(2016)}]{Counted0002}
A.~Victor Counted and Joyce~O. Arawole. 2016.
\newblock `we are connected, but constrained': internet inequality and the
  challenges of millennials in africa as actors in innovation.
\newblock \emph{Journal of Innovation and Entrepreneurship,}, 5(1).

\bibitem[{Erstad(2003)}]{Erstad001}
Tricia~L. Erstad. 2003.
\newblock Analyzing computer based patient records: a review of literature.
\newblock \emph{J Healthc Inf Manag.,}, 17(4):51--7.

\bibitem[{Forster et~al.(2008)Forster, Kyeremanteng, Hooper, Shojania, and van
  Walraven}]{Forster0001}
Alan~J. Forster, Kwadwo Kyeremanteng, Jon Hooper, Kaveh~G. Shojania, and Carl
  van Walraven. 2008.
\newblock The impact of adverse events in the intensive care unit on hospital
  mortality and length of stay.
\newblock \emph{BMC Health Services Research,}, 8:259.

\bibitem[{Fraser et~al.(2005)Fraser, Biondich, Moodley, Choi, Mamlin, and
  Szolovits}]{Fraser0002}
Hamish~S.F. Fraser, Paul Biondich, Deshen Moodley, Sharon Choi, Burke~W.
  Mamlin, and Peter Szolovits. 2005.
\newblock Implementing electronic medical record systems in developing
  countries.
\newblock \emph{Inform Prim Care,}, 13(2):83--95.

\bibitem[{Genes and Appel(2013)}]{Genes000005}
Nicholas Genes and Jacob Appel. 2013.
\newblock Ethics of data sequestration in electronic health records.
\newblock \emph{Camb Q Healthc Ethics}, 22(4):365--72.

\bibitem[{Holst et~al.(2020)Holst, Sukums, Radovanovic, Ngowi, Noll, and
  Winkler}]{Holst000002}
Christine Holst, Felix Sukums, Danica Radovanovic, Bernard Ngowi, Josef Noll,
  and Andrea~Sylvia Winkler. 2020.
\newblock Sub-saharan africa--the new breeding ground for global digital
  health.
\newblock \emph{The Lancet Digital Health,}, 2(4):160--162.

\bibitem[{Jawhari et~al.(2016)Jawhari, Ludwick, Keenan, Zakus, and
  Hayward}]{Jawhari0002}
Badeia Jawhari, Dave Ludwick, Louanne Keenan, David Zakus, and Robert Hayward.
  2016.
\newblock Benefits and challenges of emr implementations in low resource
  settings: a state-of-the-art review.
\newblock \emph{BMC Medical Informatics and Decision Making,},
  16(116):191--201.

\bibitem[{Kamadjeu et~al.(2005)Kamadjeu, Tapang, and Moluh}]{Kamadjeu0002}
Raoul~M. Kamadjeu, Euloge~M. Tapang, and Roland~N. Moluh. 2005.
\newblock Designing and implementing an electronic health record system in
  primary care practice in sub-saharan africa: a case study from cameroon.
\newblock \emph{Inform Prim Care,}, 13(3):179--86.

\bibitem[{Kavuma(2019)}]{Kavuma0001}
Michael Kavuma. 2019.
\newblock The usability of electronic medical record systems implemented in
  sub-saharan africa: A literature review of the evidence.
\newblock \emph{JMIR Hum Factors,}, 6(1):e9317.

\bibitem[{Kohli and Tan(2012)}]{Dieter0002}
Rajiv Kohli and Sharon Swee-Lin Tan. 2012.
\newblock The pen is mightier than the scalpel: the case for electronic medical
  records.
\newblock \emph{South African Journal of Industrial Engineering,},
  23(2):191--201.

\bibitem[{Kohli and Tan(2016)}]{Kohli0002}
Rajiv Kohli and Sharon Swee-Lin Tan. 2016.
\newblock Electronic health records: how can is researchers contribute to
  transforming healthcare?
\newblock \emph{MIS Quarterly,}, 40(3):553--573.

\bibitem[{Kukafka et~al.(2007)Kukafka, Ancker, Chan, Chelico, Khan, Mortoti,
  Natarajan, Presley, and Stephens}]{Kukafka01}
Rita Kukafka, Jessica~S. Ancker, Connie Chan, John Chelico, Sharib Khan,
  Selasie Mortoti, Karthik Natarajan, Kempton Presley, and Kayann Stephens.
  2007.
\newblock Redesigning electronic health record systems to support public
  health.
\newblock \emph{Journal of Biomedical Informatics,}, 40:398–409.

\bibitem[{Lakshmi et~al.(2017)Lakshmi, Gupta, and Ranjan}]{Lakshmi000005}
K.~Krithiga Lakshmi, Himanshu Gupta, and Jayanthi Ranjan. 2017.
\newblock Ussd — architecture analysis, security threats, issues and
  enhancements.
\newblock \emph{2017 International Conference on Infocom Technologies and
  Unmanned Systems (Trends and Future Directions)}.

\bibitem[{Odekunle et~al.(2017)Odekunle, Odekunle, and Shankar}]{Odekunle01}
Florence~F. Odekunle, Raphael~0. Odekunle, and Srinivasan Shankar. 2017.
\newblock Why sub-saharan africa lags in electronic health record adoption and
  possible strategies to increase its adoption in this region.
\newblock \emph{International Journal of Health Sciences Libraries,},
  11(4):59–64.

\bibitem[{Ozair et~al.(2015)Ozair, Jamshed, Sharma, and Aggarwal}]{Ozair000005}
Fouzia~F. Ozair, Nayer Jamshed, Amit Sharma, and Praveen Aggarwal. 2015.
\newblock Ethical issues in electronic health records: A general overview.
\newblock \emph{Perspect Clin Res.}, 6(2):73--76.

\bibitem[{Rodríguez-Castelán et~al.(2021)Rodríguez-Castelán, Ochoa, Lach,
  and Masaki}]{Castelan0002}
Carlos Rodríguez-Castelán, Rogelio~Granguillhome Ochoa, Samantha Lach, and
  Takaaki Masaki. 2021.
\newblock Mobile internet adoption in west africa.
\newblock \emph{World Bank Group: Poverty and Equity Global Practice, Policy
  Research Working Paper 9560}.

\bibitem[{Sabi et~al.(2018)Sabi, Uzoka, Langmia, Njeh, and Tsuma}]{Sabi0002}
Humphrey~M. Sabi, Faith-Michael~E. Uzoka, Kehbuma Langmia, Felix~N. Njeh, and
  Clive~K. Tsuma. 2018.
\newblock A cross-country model of contextual factors impacting cloud computing
  adoption at universities in sub-saharan africa.
\newblock \emph{Information Systems Frontiers,}, 20:1381--1404.

\bibitem[{Schmitt and Wofford(2002)}]{Schmitt001}
Karl~F. Schmitt and David~A. Wofford. 2002.
\newblock Financial analysis projects clear returns from electronic medical
  records.
\newblock \emph{Healthc Financ Manage.,}, 56(1):52--7.

\bibitem[{Schwartz et~al.(2019)Schwartz, Gao, Geng, Mody, Mikhail, and
  Cho}]{Schwartz1000005}
John~T. Schwartz, Michael Gao, Eric~A. Geng, Kush~S. Mody, Christopher~M.
  Mikhail, and Samuel~K. Cho. 2019.
\newblock Applications of machine learning using electronic medical records in
  spine surgery.
\newblock \emph{Neurospine.}, 16(4):643--653.

\bibitem[{Sood et~al.(2008)Sood, Nwabueze, Mbarika, Prakash, Chatterjee, Ray,
  and Mishra}]{Sood0002}
Sanjay~P. Sood, Stacie~N. Nwabueze, Victor~W.A. Mbarika, Nupur Prakash, Samir
  Chatterjee, Pradeep Ray, and Saroj Mishra. 2008.
\newblock Electronic medical records: A review comparing the challenges in
  developed and developing countries.
\newblock \emph{In Proc. of the 41st Hawaii International Conference on System
  Sciences}.

\bibitem[{Stiell et~al.(2003)Stiell, Forster, Stiell, and van
  Walraven}]{Stiell01}
Andrew Stiell, Alan~J. Forster, Ian~G. Stiell, and Carl van Walraven. 2003.
\newblock Prevalence of information gaps in the emergency department and the
  effect on patient outcomes.
\newblock \emph{CMAJ.,}, 169(10):1023--8.

\bibitem[{Syzdykova et~al.(2017)Syzdykova, Malta, Zolfo, Diro, and
  Oliveira}]{Syzdykova0002}
Assel Syzdykova, André Malta, Maria Zolfo, Ermias Diro, and José~Luis
  Oliveira. 2017.
\newblock Open-source electronic health record systems for low-resource
  settings: Systematic review.
\newblock \emph{JMIR Med Inform,}, 5(4):e44.

\bibitem[{Wong et~al.(2018)Wong, Horwitz, Zhou, and Toh}]{Wong333000005}
Jenna Wong, Mara~Murray Horwitz, Li~Zhou, and Sengwee Toh. 2018.
\newblock Using machine learning to identify health outcomes from electronic
  health record data.
\newblock \emph{Curr Epidemiol Rep}, 5(4):331--342.

\bibitem[{Yang et~al.(2022)Yang, Mu, Peng, Li, Wang, Wang, Wang, and
  Han}]{Yang1000005}
Xinyu Yang, Dongmei Mu, Hao Peng, Hua Li, Ying Wang, Ping Wang, Yue Wang, and
  Siqi Han. 2022.
\newblock Research and application of artificial intelligence based on
  electronic health records of patients with cancer: Systematic review.
\newblock \emph{JMIR Med Inform}, 10(4):e33799.

\bibitem[{Zhou et~al.(2015)Zhou, Herselman, and Coleman}]{Zhou000005}
Munyaradzi Zhou, Marlien Herselman, and Alfred Coleman. 2015.
\newblock Ussd technology a low cost asset in complementing public health
  workers' work processes.
\newblock \emph{BioInformatics and Bio Engineering Conference,}, pages 57--64.

\end{thebibliography}
\bibliographystyle{acl_natbib}




\end{document}